\documentstyle[prl,aps,epsf,psfig,multicol,amssymb]{revtex}

\def\N{{\mathbb N}}

\begin{document}
\draft

\title{Spontaneous plaquette formation in the SU(4) Spin-Orbital ladder}
\author{Mathias van den Bossche$^1$, 
Patrick Azaria$^2$, Philippe Lecheminant$^3$ and Fr\'ed\'eric Mila$^1$}
\address{
$^1$ Laboratoire de Physique Quantique -- Universit\'e Paul Sabatier -- 31062 Toulouse Cedex 04 - France \\
$^2$ Laboratoire de Physique Th\'eorique des Liquides -- Universit\'e Pierre et Marie Curie -- 75252 Paris Cedex 05 - France \\
$^3$ Laboratoire de Physique Th\'eorique et Mod\'elisation -- Universit\'e de Cergy-Pontoise -- 95301 Cergy-Pontoise Cedex -- France
}
\maketitle
\begin{abstract}
The low-energy properties of the SU(4) spin-orbital model on a two-leg
ladder are studied by a variety of analytical and numerical techniques.
Like in the case of SU(2) models, there is a singlet-multiplet gap
in the spectrum, but the ground-state is two-fold degenerate. An interpretation
in terms of SU(4)-singlet plaquettes is proposed. The implications for general
two-dimensional lattices are outlined.
\end{abstract}
\pacs{PACS numbers: 75.10.Jm, 11.30.-j, 75.40.Mg}
\begin{multicols}{2}
\narrowtext

The properties of Mott insulators with orbital degeneracy is attracting a lot
of attention with the increasing evidence that this degeneracy can have many
other consequences apart from the standard cooperative Jahn-Teller effect. One of the
possibilities that seems to be realized in LiNiO$_2$\cite{kitaoka} 
is that the additional orbital
degree of freedom prevents the system from ordering in both the orbital and spin
channels. It was suggested in a recent paper by Li {\it et al.}\cite{li1} 
that this might
occur if spin and orbital degrees of freedom play a very symmetric role, like in
the SU(4) symmetric version of the Kugel'-Khomski\u\i ~model\cite{kugel} 
defined by 
the Hamiltonian:
\begin{equation}
H=\sum_{ij} J_{ij} (2\vec s_i.\vec s_j +\frac{1}{2})
(2\vec \tau_i.\vec \tau_j +\frac{1}{2})
\end{equation}
Such a Hamiltonian is indeed a good starting point for LiNiO$_2$ due to the local
symmetry and the strong Hund's rule coupling, but its properties are only beginning
to be understood. The fundamental difference with SU(2) models stems
from the fact that 
it takes at least 4 sites to make an SU(4) singlet. 
For the 1D version of the model,
which is fairly well understood both at zero\cite{sutherland,affleck,ueda,azaria1} 
and finite temperature\cite{frischmuth,mila}, this shows up
as a four-site periodicity of the correlation function. In 2D lattices, it was 
argued by Li {\it et al.}\cite{li1,li2} that the system might prefer to 
form local SU(4) singlet
plaquettes in the ground state rather than developing long-range order. While exact 
diagonalizations (ED) of the model on a square lattice indeed support this 
conjecture\cite{vdb}, the lack of analytical results in any limit prevents one from
drawing definite conclusions.

\begin{figure}
\centerline{\psfig{file=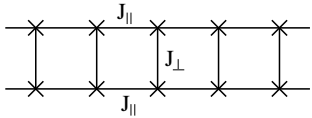,height=1.5cm,angle=0}}
\caption{The spin-orbital ladder.}
\label{ladder}
\end{figure}

In this Letter, we have adopted another strategy and decided to study the simplest 
lattice in which plaquettes 
might form, namely the two-leg ladder (Fig. 1). As we shall see, exact diagonalizations
suggest that the ground state is a two-fold degenerate plaquette solid with gapped
multiplet excitations. The important step forward though is that analytical 
results can be 
obtained in both the weak and strong rung limits finally putting this plaquette 
picture on very firm grounds.

The SU(4) spin-orbital model on a ladder is defined by the Hamiltonian  
\begin{eqnarray}
H = &J_\parallel \sum_{i,\alpha} 
(2{\vec s}_{i,\alpha}.{\vec s}_{i+1,\alpha}+\frac{1}{2}) 
(2{\vec \tau}_{i,\alpha}.{\vec \tau}_{i+1,\alpha}+\frac{1}{2})\nonumber\\
+   &J_\perp \sum_{i} (2{\vec s}_{i,1}.{\vec s}_{i,2}+\frac{1}{2}) 
(2{\vec \tau}_{i,1}.{\vec \tau}_{i,2}+\frac{1}{2}),
\label{hamilt}
\end{eqnarray}
where a site on the two-leg ladder is described by its rung number $i$ 
and its chain index $\alpha=1,2$, ${\vec s}_{i,\alpha}$ is a spin one-half 
operator at site $(i,\alpha)$ and  ${\vec \tau}_{i,\alpha}$ is an isospin 
one-half corresponding to the orbital degree of freedom on the same site
(see Fig. \ref{ladder}).

\begin{figure}
\centerline{\psfig{file=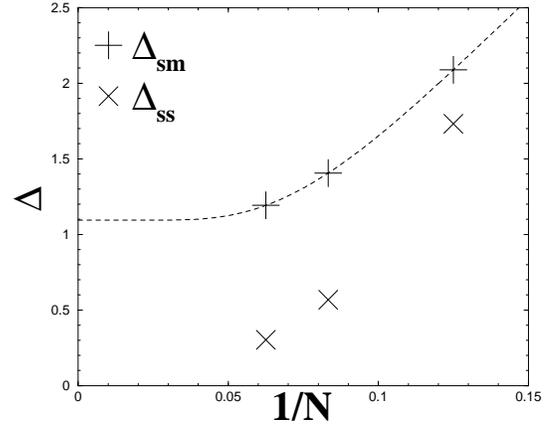,height=6cm,angle=-90}}
\caption{Scaling of the singlet-multiplet gap (+) and the 
singlet-singlet gap (X). The dashed line is a fit with
$\Delta_{sm}=\Delta_{sm}^\infty+Be^{-N/\xi}$. It is clearly
consistent with a non-zero value of $\Delta_{sm}$ in the
$N\rightarrow\infty$ limit, whereas the behavior of $\Delta_{ss}$ 
is consistent with a 
zero value in the same limit.} 
\label{scalgaps}
\end{figure}

{\sc Isotropic ladder --}
We start by considering the case that is closer to 2D, namely the isotropic 
limit $J_\perp=J_\parallel=J$.
Taking advantage of all symmetries (translation, rung-parity and 
SU(4) quantum numbers $s^z_{\rm tot}$, $\tau^z_{\rm tot}$ 
and $s\tau^z_{\rm tot}=\sum_i s^z_i\tau^z_i$), we have obtained the low-energy
spectrum on clusters with 8, 12 and 16 sites 
with Lanczos ED using periodic boundary conditions in the chain direction.
The results can be summarized as follows. For all clusters, the ground state
is an SU(4) singlet, and the first multiplet excitation is at relatively 
high energy. Besides, a plot of this energy gap $\Delta_{\rm sm}$ as a function
of $1/N$ (see Fig. 2) strongly suggests that this gap remains in the 
thermodynamic limit. In addition to this multiplet excitation there is always 
one low-lying singlet inside this gap (2 in the special case of N=8). The 
splitting between this excited singlet and the ground state $\Delta_{\rm ss}$
is plotted in Fig. \ref{scalgaps} as a function of $1/N$. Although it is difficult to draw
definite conclusions with only 3 sizes, these results strongly suggest that this
gap vanishes, and that the ground state is two-fold degenerate in the
thermodynamic limit.

\begin{figure}
\centerline{\psfig{file=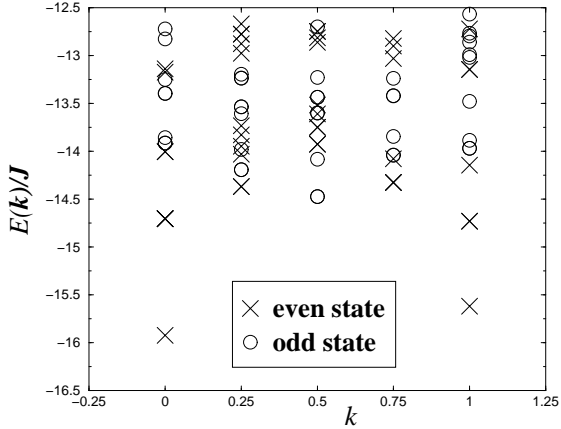,height=6cm,angle=-90}}
\caption{ Dispersion of the low-lying states in the $N=16$ site ladder. 
The two lowest-lying states with $k=0$ and $k=\pi$ are $SU(4)$ singlets.
There is a local minimum at $k=\pi/2$.
}
\label{disp} 
\end{figure}

Fig. \ref{disp} shows the dispersion of the low-lying states for
16 sites. Two important facts are to be noticed here. First, 
the ground state and the next singlet lie in the $k=0$ and 
$k=\pi$ sectors respectively. Second, the dispersion has a local 
minimum at $k=\pi/2$,
which announces the soft mode at $k=\pi/2$ found in the chain limit where 
$J_\perp=0$\cite{sutherland,affleck,ueda,frischmuth,li2}.

All these results can be qualitatively interpretated in terms of plaquette
coverings of the ladder \cite{li1,li2,vdb}. The plaquette is the ground state 
of the four-site SU(4) spin-orbital system. It is the smallest SU(4)
singlet one can build with a system of degrees of freedom in the fundamental 
($d=4$) representation of SU(4). It undergoes extremely strong fluctuations 
which minimize the energy by link to $E_0/N=-J$. It is thus a very stable 
object and can be used to describe the physics of many realizations of this
model.  For a larger system with $N=4p,\;p\in\N$ , one can build tensor-product 
states as coverings by such plaquettes in which the system is `tetramerized'. 
\begin{figure}
\centerline{\psfig{file=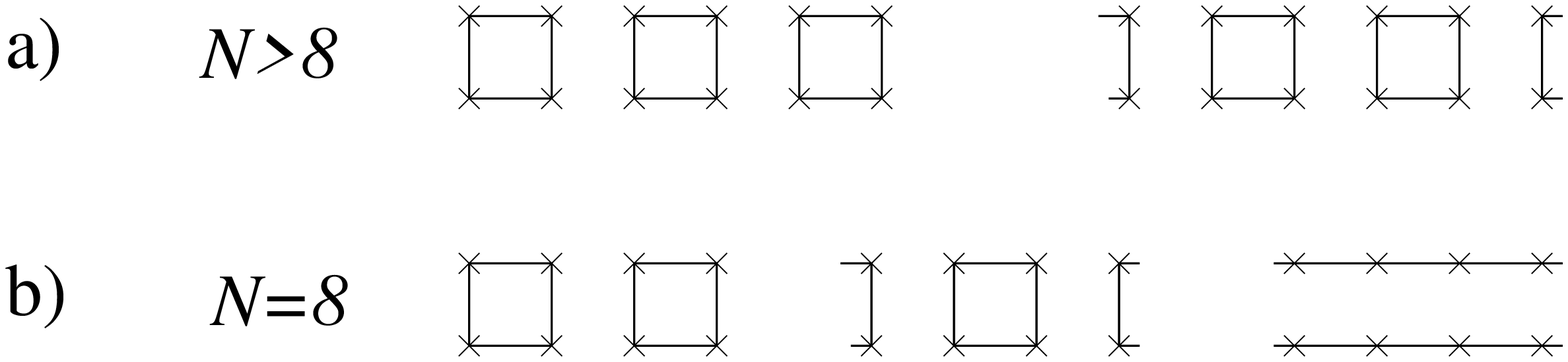,height=2cm,angle=0}}
\caption{Plaquette coverings of the $SU(4)$ ladder. }
\label{plaqcov}
\end{figure}
In the case of the two-chain ladder, the number of such plaquette coverings 
is two, as shown in Fig. \ref{plaqcov}a. These two coverings differ by a 
translation by one lattice spacing along the direction of the ladder. They 
are rung-symmetric, thus having a $+$ rung-parity. A symmetric and an 
antisymmetric linear combination of these states can be built, giving one $k=0$ and
one $k=\pi$, $+$ rung-parity SU(4) singlet state. 
These quantum numbers agree with our ED results.
In the 
special case of the $N=8$ ladder, which has the same topology as a cube, 
there is an extra covering, corresponding to the third pair of cube faces 
that can be occupied by each of the two plaquettes (see Fig. \ref{plaqcov}b). 
So the plaquette picture predicts 3 low-lying states in this special case,
again in agreement with our ED results.
Besides, if the ground state is a product of singlet plaquettes, multiplet
excitations require the breaking of a plaquette, with a finite energy cost 
equal to $2J$ minus a correction due to the delocalization of this defect,
again in agreement with our ED results.

So all the basic features of our numerical results are qualitatively 
reproduced by this simple plaquette picture.
The following strong coupling approach 
provides more elements to support this tetramerization picture.

{\sc Strong coupling --}
We now turn to the strong rung limit $J_\perp\gg J_\parallel$. When
$J_\parallel=0$, the ground state is obtained by putting each rung in one
of its 6 ground states. The ground state of a rung can be thought of as
the 6-dimensional irreducible representation of SU(4), or as the set
of states (spin singlet $\times$ orbital triplet) and (spin triplet 
$\times$ orbital singlet). To perform strong coupling analysis, we
thus have to determine the effective Hamiltonian that will lift the
degeneracy in this $6^{N_{\rm rung}}$-fold degenerate subspace.

To first order in $J_\parallel$, we need only to consider the coupling between
two adjacent rungs. Denoting by (12) and (34) the sites of two adjacent
rungs, we can actually couple them in two equivalent ways: 1 to 3 and 2
to 4 ($H_1$) or 1 to 4 and 2 to 3 ($H_2$). To first order, the effective
Hamiltonians corresponding to $H_1$ and $H_2$ can be formally written:
\begin{equation}
H^{\rm eff}_{1,2}=\sum_{i,j} |i\rangle V_{ij}^{1,2} \langle j|
\label{heff}
\end{equation}
where the sum over $i,j$ runs over the 36 states of the $J_\parallel=0$ limit.
Now, to go from $H^{\rm eff}_1$ to $H^{\rm eff}_2$, we just have to exchange
sites 3 and 4. But this transforms any ket $|i\rangle$ (respectively 
bra $\langle j|$) 
in the sum of Eq. (\ref{heff}) into $-|i\rangle$ (respectively $-\langle j|$)
since this permutation just changes the sign of the singlet 
and leaves the triplet
invariant. Given the form of the effective Hamiltonian, we thus have 
$H^{\rm eff}_1=H^{\rm eff}_2$.

Now the sum of these Hamiltonians $H_0=H_1+H_2$ is a very simple operator because
each site is coupled to both sites of the opposite rung. In terms of the 
15-dimensional vector $\vec A$ of each rung, whose components are 
the generators 
of SU(4),
it can be written
\begin{equation}
H_0=\frac{J_\parallel}{4}[\vec A_{12}.\vec A_{34}]+J_\parallel
\end{equation}
with $\vec A_{12}=\vec A_1+\vec A_2$ and $\vec A_{34}=\vec A_3+\vec A_4$.
As shown in Ref. \cite{vdb}, this Hamiltonian can be rewritten in terms of Casimir
operators as:
\begin{equation}
H_{0}= 4J_\parallel C_{1234} -4J_\parallel(C_{12}+C_{34})+J_\parallel
\end{equation}
So the spectrum obtained when coupling two irreducible representations of 
dimension 6 consists of three levels with degeneracy 1, 15, 20 and with energy
$-4 (J_\parallel+J_\perp)$, $-4 J_\parallel$ and 
$-4 (J_\parallel-J_\perp)$ respectively.
The spectrum of $H_0$ is thus linear in $J_\parallel$. So $H^{\rm eff}_0=H_0$, 
and since
$H^{\rm eff}_1=H^{\rm eff}_2$ and $H_1+H_2=H_0$, we reach the conclusion that 
$H^{\rm eff}_1=\frac{1}{2}H_0$.

\begin{figure}
\centerline{\psfig{file=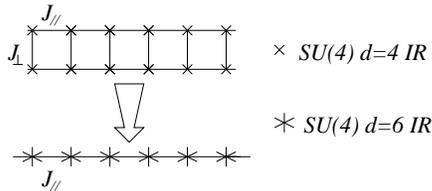,height=2.5cm,angle=0}}
\caption{(a) The strong-coupling low-energy effective hamiltonian for the
$SU(4)$ spin-orbital ladder. ($\times$) stand for $d=4$ IR degrees of freedom
while ($\ast$) stands for $d=6$ IR degrees of freedom.}
\label{effchain}
\end{figure}

A more pedestrian way to reach this conclusion consists in calculating the spectrum of
$H_1$. For small $J_\parallel$, the levels are linear in $J_\parallel$, and 
we have checked numerically that the splittings and degeneracies of $H_1$
correspond to $H_0/2$ when $J_\parallel$ is small.
Back to the ladder, the first-order Hamiltonian thus writes up to a constant
\begin{equation}
H^{\rm eff} = \frac{J_\perp}{8} \sum_{i,j} \vec A_{i,\rm tot}\vec A_{j,\rm tot}
\end{equation}
where $\vec A_{i,\rm tot}=\vec A_{i1}+\vec A_{i2}$.  $H_0$ is thus 
nothing but the SU(4) Hamiltonian for the $\vec A_{i,\rm tot}$ rung degree
of freedom. In other words, the effective Hamiltonian is the 1D SU(4) model 
in the antisymmetric 6-dimensional representation (see Fig. \ref{effchain}).
This situtation is analogous
to going from the $S=1/2$ SU(2) spin ladder with ferromagnetic rungs
to the $S=1$ SU(2) spin chain.

This simple form of the effective Hamiltonian has very interesting consequences.
The $d=6$ representation of SU(4) is a self-conjugate and antisymmetric
representation of an SU(N) group with $N$ even.  It thus falls in the 
cases where the Lieb-Schulz-Mattis-Affleck theorem \cite{lieb,affleck1}
states that the SU(4) Hamiltonian should either have a non-degenerate 
ground-state followed by gapless excitations or have a degenerate ground-state.
Affleck, Arovas, Marston and Rabson \cite{affleck2} have shown that the 
ground-state is a two-fold degenerate singlet, and breaks translation 
invariance. More precisely, the two ground-states 
are spontaneously dimerized, nearest-neighboring sites forming $SU(4)$
singlets either between neighbors $(2n,2n+1)$ or  between neighbors 
$(2n+1,2n+2)$. Above these ground states there is a gap to magnon-like or 
soliton-like excitations\cite{marston,affleck2}. 
DMRG calculations \cite{onufriev} 
confirmed this picture of dimer order with short-range spin-spin correlations 
($\vec A$-spin correlation length of the order of the lattice spacing).

This strong-coupling regime is very similar to the physics we have 
characterized numerically in the intermediate coupling regime and strongly
supports our plaquette interpretation. 
First of all, the spectrum has the same properties in both cases:
The two plaquette-states break translation 
symmetry, have a short correlation length, and the first excitation 
has to be built breaking one plaquette, thus leading to a gap. 
Besides, a tetramerization of a ladder is equivalent to a dimerization
in terms of rungs.
We now come to the 
weak-coupling regime to show how the situation sets in when two SU(4)
gapless chains are coupled to form a ladder.

{\sc Weak coupling --}
The weak coupling approach proceeds in an analogous way as in the
SU(2) ladder. In the absence of interchain coupling ($J_{\perp} =0$)
the Hamiltonian (1) describes two decoupled SU(4) spin chains and
is exactly solvable by the Bethe ansatz\cite{sutherland}. The system is
gapless and the low energy physics is described by six (three for each chain)
massless bosons and is controlled by the fixed point Hamiltonians
of two decoupled Wess-Zumino-Novikov-Witten (WZNW) SU(4)$_1$
models\cite{affleck} with central charge $c= 3 + 3 =6$. As in the SU(2)
ladder, the
strategy to tackle with the weak coupling regime is to look at the stability
of the infrared fixed point with respect to the interchain coupling.
To this end one needs the low energy expressions for the SU(4)
spin densities in terms of the WZNW fields which has been obtained
in Refs.\cite{affleck,assaraf}
\begin{equation}
{\cal S}_a^A \simeq {\cal J}_{aR}^A + {\cal J}_{aL}^A 
+ \left[e^{i\pi x/2a_0}  {\cal N}_a^A + H.c.
\right] + (-1)^{x/a_0} { n}_a^A
\label{spindensity}
\end{equation}
where ${\cal S}_a^A$ are the 15 SU(4) spin densities of chain
index  $a=1,2$ with components ${\cal J}_{a R,L}^A$ at $k=0$,
${\cal N}_a^A$ at $2k_F=\pi /2a_0$, and ${ n}_a^A$ at $4k_F=\pi /a_0$.
The uniform part of the spin density is the SU(4) spin current
with scaling dimension $d_0 = 1$. The other oscillating parts, ${\cal
N}_a^A$ and ${ n}_a^A$ are WZNW primary fields with scaling dimensions 
$d_{ 2k_F}= 3/4$ and $d_{ 4k_F}= 1$ and transform respectively
into the fundamental and the antisymmetric two-rank tensorial 
representations of SU(4). With these results, one can obtain the low energy
effective Hamiltonian of the weakly coupled SU(4) spin ladder
\begin{eqnarray}
{\cal H}_{eff} = \frac{2\pi v}{5} \sum_{a=1}^{2}
\left( {\cal  J}_{aR}^A {\cal  J}_{aR}^A + (R \rightarrow L) \right)
\nonumber \\
+ J_{\perp} \left( {\cal  J}_{1R}^A {\cal  J}_{2L}^A 
+ {\cal  J}_{2R}^A {\cal  J}_{1L}^A + { n}_1^A { n}_2^A \right)
\nonumber \\
+  J_{\perp} \left( {\cal N}_1^A {\cal N}_2^{A \dagger} + {\cal N}_2^A {\cal N}_1^{A \dagger}
\right),
\label{heffectif}
\end{eqnarray}
where we have dropped as usual the marginally irrelevant current-current
in-chain interactions as well as the interaction between the current
of the two chains with the same chirality that renormalizes the spin velocity.
The interacting part of Eq. (\ref{heffectif}) has two contributions.
One comes from the uniform and $4k_F$ parts
of the spin densities (\ref{spindensity}). It is marginal with scaling
dimension $2$. The other contribution, which stems from  the $2k_F$ spin
 densities,
 is a strongly relevant perturbation with scaling dimension $3/2$ and thus
governs the low energy behavior of the model. As an immediate
consequence  we conclude that a gap $\Delta \sim J_{\perp}^2$ opens in the
spectrum. The delicate point however is whether or not some gapless modes
 survives in the infrared. This issue can be investigated by means of the
Abelian bosonization of the SU(4) spin densities (\ref{spindensity}).
Using the results of Ref.\cite{assaraf} we have expressed the effective
Hamiltonian (\ref{heffectif}) in terms of the six bosonic fields
that describe the ultraviolet fixed point. The resulting bosonized
Hamiltonian is too lengthly to be reproduced here but  it can be
shown that all degrees of freedom are massive.

At this point, the physically relevant question is whether the ``plaquette''
picture drawn from the strong coupling analysis survives at weak coupling, and
in particular whether the nature of the low lying excitations at strong
coupling changes as the interchain coupling is reduced. In the SU(2) ladder,
the nature of the low energy spectrum is the same in both limits and is 
captured by the weak coupling approach. In this case it has
been shown in Ref.\cite{shelton} that the effective Hamiltonian separates 
into two decoupled free field  theories (free massive real fermions)
that describe both singlet and triplet  sectors. 
In contrast in the SU(4) model an analogous decomposition does not hold.
Indeed, the leading part of the bosonized Hamiltonian does not split
into two parts that account for the six-dimensional (antisymmetric)
and tenth-dimensional (symmetric) SU(4) irreducible representation:  
All degrees of freedom
strongly interact. 
In the simplest hypothesis, we expect that no phase transition occurs
between weak and strong couplings but rather a smooth cross-over 
to the plaquette picture described above. 

{\sc Conclusion --}
In summary, coming back to the issue raised in the introduction, we now have 
definite evidence that the presence of an SU(4) symmetry can indeed have
very dramatic consequences for lattices in which plaquettes can form.  
In the case of the two-leg ladder we have shown that there is a spontaneous plaquette
formation that leads to a degenerate singlet ground state in an otherwise gapped
spectrum. This leads naturally to the conjecture that, for more general
lattices, there will be low lying singlets, and that such plaquette
coverings provide a good variational basis to describe them. Work is in
progress along these lines.

We acknowledge very useful discussions with Karlo Penc and Fu-Chun Zhang.
The numerical computations were performed on the Cray supercomputers of the IDRIS
(Orsay, France).

\end{multicols}

\begin{thebibliography}{0}

\bibitem{kitaoka} Y. Kitaoka, T. Kobayashi, A.K\={o}da, H. Wakabayashi, 
Y. Niino, H. Yamakage, S. Taguchi, K. Amaya, K. Yamamura, M. Takano, 
A. Hirano, and R. Kanno, J. Phys. Soc. Jpn. {\bf 67}, 3703 (1998).

\bibitem{li1} Y. Q. Li, M. Ma, D. N. Shi, and F.-C. Zhang, Phys. Rev. Lett. {\bf 81}, 3527 (1998).

\bibitem{kugel} For a review of the standard Jahn-Teller decoupled
spin-orbital systems, see K. I. Kugel' and D.I. Khomski\u \i ~ Usp. Fiz. Nauk. 
{\bf 136} 631 (1982), Sov. Phys. Usp. {\bf 25}, 231 (1982).

\bibitem{sutherland} B. Sutherland, Phys. Rev. B {\bf 12}, 3795 (1975).

\bibitem{affleck} I. Affleck, Nucl. Phys. {\bf B265}, 409 (1986);
Nucl. Phys. {\bf B305}, 582 (1988).

\bibitem{ueda} Y. Yamashita, N. Shibata, and K. Ueda, Phys. Rev. B 
{\bf 58}, 9114 (1998).

\bibitem{azaria1} P. Azaria, A. O. Gogolin, P. Lecheminant,
and A. A. Nersesyan, Phys. Rev. Lett. {\bf 83}, 624 (1999).

\bibitem{frischmuth} B. Frischmuth, F. Mila, and M. Troyer,
Phys. Rev. Lett. {\bf 82}, 835 (1999).

\bibitem{mila} F. Mila, B. Frischmuth, A. Deppeler, and M. Troyer,
Phys. Rev. Lett. {\bf 82}, 3697 (1999).

\bibitem{li2} Y.-Q. Li, M. Ma, D.-N. Shi, and F.-C. Zhang, Phys. Rev. B {\bf 60}, 12781 (1999).  

\bibitem{vdb} M. van den Bossche, F. Mila, and  F.-C. Zhang, cond-mat/0001051 preprint.

\bibitem{lieb} E.H. Lieb, T.Schultz, and D.J. Mattis, Ann. Phys. (N.Y.) {\bf 16}, 407 (1961).

\bibitem{affleck1} I. Affleck and E. Lieb, Lett. Math. Phys. {\bf 12}, 57 (1986).

\bibitem{affleck2} I. Affleck, D.P. Arovas, J.B. Marston, and D.A. Rabson, Nucl. Phys. B {\bf 366}, 467 (1991).  

\bibitem{marston} J.B. Marston and I. Affleck, Phys.  Rev. B {\bf 39}, 11538 (1989).  

\bibitem{onufriev} V.A. Onufriev and J.B. Marston Phys. Rev. B {\bf 59}, 12573 (1999).

\bibitem{assaraf}
R. Assaraf,  P. Azaria, M. Caffarel, and P. Lecheminant, Phys. Rev. B {\bf
60}, 2299 (1999).

\bibitem{shelton} D. G. Shelton, A. A. Nersesyan, and A. M. Tsvelik,
Phys. Rev. B {\bf 53}, 8521 (1996).


\end{thebibliography}
\end{document}